\documentclass[12pt,showpacs,preprintnumbers,amsmath,aps,amssymb]{revtex4}
\usepackage{graphicx}
\usepackage{dcolumn}
\usepackage{bm}
\begin{document}

\title{(Non)commutative isotropization in Bianchi I with Barotropic perfect fluid and $\Lambda$ Cosmological}
\author{C. Ortiz$^{1}$}
\email{ortizgca@fisica.ugto.mx}
\author{E. Mena-Barboza$^{1}$}
\email{emena@fisica.ugto.mx}
\author{M. Sabido$^{1}$}
\email{msabido@fisica.ugto.mx}
\author{J. Socorro$^{1,2}$.}
\email{socorro@fisica.ugto.mx}
\affiliation{$^1$ Instituto de F\'{\i}sica de la Universidad de Guanajuato,\\
 A.P. E-143, C.P. 37150, Le\'on, Guanajuato, M\'exico\\
 $^2$ Facultad de Ciencias de la Universidad Aut\'onoma del Estado de M\'exico,\\
Instituto Literario  No. 100, Toluca, C.P. 50000, Edo de Mex, M\'exico}%

\date{\today}

\begin{abstract}
We present the classical solutions to the Einstein field
equations derived using the WKB-like and Hamilton procedures.
The investigation is carried out in the commutative and
noncommutative scenario for the  Bianchi type I cosmological model
coupled to barotropic perfect fluid and $\lambda $ Cosmological for two
different gauges. Noncommutativity is achieved by modifying the symplectic
structure considering that  all minisuperspace  variables $\rm q^i$ does not
commute and by a deformation between all the minisuperspace variables. In the gauge
N=1, it is possible to obtain that the anisotropic parameter $\rm \beta_{\pm nc}$
 tend to a  constant curvature for large period of time considering different
 values in the noncommutative parameters $\theta$ and cosmological term. However,
 this  behavior give the idea that is necessary introduce other class of matter in the
 models, for to have a real isotropization in the model, such as dark energy or dark
 matter.
\end{abstract}

\maketitle

\section{Introduction}
Recently, a great  interest has been generated in noncommutative
spacetimes \cite{connes,varilly,douglas}, mainly due to the fact
that there are of strong motivations in the development of string
and M-theories \cite{banks,seiberg}. A different approach to
noncommutativity is through the introduction of noncommutative
fields \cite{carmona1}, that is, fields of their conjugate momenta
are taken as noncommuting. There are several approaches in
considering the notion of noncommutativity in cosmology, that could
be the best alternative in the absence of a comprehensive and
satisfactory theory from string theory. This analysis has been
studied in many works \cite{lizzi,branden}. Here, taking coordinates
as noncommuting, it  has been shown that noncommutativity affects
the spectrum of Cosmic Microwave Background. For example, in
\cite{lizzi}, noncommutative geometry suggest a non local inflaton
field that changes the gaussianity and isotropy property of
fluctuations. In cosmological systems, since the scale factors,
matter fields and their conjugate momenta play the role of dynamical
variables of the system, introduction of noncommutativity by
adopting the noncommutativity between all fields, is particularly
relevant. The simplest noncommutative classical and quantum
cosmology of such models have been studied in different works
\cite{compean,barbosa1,barbosa2,mena}.

On the other hand,  there is a renewed interest on noncommutative theories to explain the
appropriate modification of classical General Relativity, and hence of spacetime symmetries
at short-distance scales, that implies modifications at large scales.
 General quantum mechanics arguments indicate that, it is not possible to measure a classical
background spacetime at the Planck scale, due to the effects of
gravitational backreaction \cite{Szabo1}. It is therefore tempting
to incorporate the dynamical features of spacetime at deeper
kinematical level using the standard techniques of noncommutative
classical field theory based in the so called Moyal product in which
for all calculations purposes (differentiation, integration, etc.) the space
time coordinates are treated as ordinary (commutative) variables and noncommutativity
enters into play in the way in which fields are multiplied \cite{Szabo2}.
Using a modified symplectic structure  on the space variables in the Hamilton approach,
as we are trying with the idea of noncommutative space time, we propose that the
minisuperspace variables do not commute, for that purpose we will modified the Poisson
structure,
this approach does not modify the hamiltonian structure in the noncommutative fields.
In the approach used, we choose that the momentas in both spaces, are the same,
$\rm P_{q^\mu_{nc}}= P_{q^\mu}$, it is say, they commute in both spaces.

Another way to extract useful dynamical information is through
the WKB semiclassical approximation to the quantum Wheeler-DeWitt equation using
the wave function $\rm \Psi=e^{iS(q^\mu)}$.
In this sense, we consider the usual approximation in the derivatives and
the corresponding relation between the Einstein-Hamilton-Jacobi (EHJ) equation,
it was possible to obtain classical solutions at the master equation
found by this procedures. The classical field equations were checked for all solutions,
using  the REDUCE 3.8 algebraic packages.

The main idea in this paper is to obtain both, the commutative $(\rm \Omega,\beta_\pm)$
and noncommutative $(\rm \Omega_{nc},\beta_{\pm nc})$ classical solution of the
Einstein field equation in General Relativity for the Bianchi Class A models,
without solve these field equations, using two alternative approaches, known as
WKB semiclassical approximation and Hamilton approach.
Using these solutions in the gauge N=1 (the physical gauge), we can inffer
if the anysotropic parameters $\beta_{\pm nc}$ suffers changes toward isotropic ones
(a constant or zero value). This analysis is considered in particular  with the Bianchi
type I, coupled to barotropic perfect fluid and cosmological term. In this case, we can observe that when the
cosmological constant decrease in its value, the isotropization is more notorious for a larger period time.

The paper is then organized as follows. In section II, we obtain  the WDW equation
including the barotropic matter contribution, and the corresponding commutative
classical  solutions for the cosmological  Bianchi type I,  in the gauge $N=1$,
by the WKB semiclassical
 approximation and Hamilton procedure. Section III is devoted to the noncommutative
 classical solutions and the analysis of the isotropization is made too, in the physical gauge N=1.
 Final remarks are presented in Section IV. For completeness,
  we can follow a similar prescription for the gauge $\rm N=24 e^{3\Omega}$, where the noncommutative cosmological model
  is always anisotropic. So, we  present the corresponding solutions for both scenarios, the commutative,
  appendix A and noncommutative, appendix B.

Let us begin by recalling canonical formulation of the ADM formalism to the
diagonal Bianchi Class A cosmological models. The metrics have the form
\begin{equation}
\rm ds^2= -(N^2- N^j N_j)dt^2 + e^{2\Omega(t)} e^{2\beta_{ij}(t)} \, \omega^i \omega^j,
\label{metrica}
\end{equation}
where $\rm N$ and $\rm N_i$ are the lapse and shift functions, respectively, $\Omega(t)$ is a scalar and
$\rm \beta_{ij}(t)$ a 3x3 diagonal matrix,
$\rm \beta_{ij}=diag(\beta_+ +\sqrt{3} \beta_-,\beta_+ -\sqrt{3} \beta_-, -2\beta_+)$,
$\rm \omega^i$ are one-forms that  characterize  each cosmological Bianchi
type model, and that obey $\rm d\omega^i= \frac{1}{2} C^i_{jk} \omega^j \wedge \omega^k,$
$\rm C^i_{jk}$ the structure constants of the corresponding invariance group \cite{ryan}.
The metric for the Bianchi type I, takes the form
\begin{equation}
\rm ds^2_I=  - N^2dt^2 + e^{2\Omega} e^{2\beta_+ +2\sqrt{3}\beta_-} dx^2
+e^{2\Omega} e^{2\beta_+ -2\sqrt{3}\beta_-} dy^2 + e^{2\Omega} e^{-4\beta_+ } dz^2, \label{bi}
\end{equation}
The corresponding lagrangian density is
\begin{equation}
\rm L_{Total}=\sqrt{-g}\, (R-2\Lambda)+ L_{matter},
\label{lagra}
\end{equation}
and using (\ref{bi}), this have the following form
\begin{equation}
 \rm L= \rm  6 e^{3\Omega} \left[-\frac{\dot \Omega^2}{N}
 +\frac{\dot \beta_+^2}{N} +\frac{\dot \beta_-^2}{N}
  -\frac{\Lambda}{3} N + \frac{8}{3} \pi G N \rho
 \right].
 \end{equation}
 where the overdot denotes time derivatives. The canonical momentas to coordinate fields are defined in the usual way
  \begin{eqnarray}
 \rm P_\Omega=\rm \frac{\partial L}{\partial \dot \Omega}= -12 e^{3\Omega}\frac{\dot \Omega}{N}, \ \ \ \
\rm P_+=\rm \frac{\partial L}{\partial \dot \beta_+}= 12 e^{3\Omega}\frac{\dot \beta_+}{N}, \ \ \ \ \rm P_-=\rm \frac{\partial L}{\partial \dot \beta_-}= 12 e^{3\Omega}\frac{\dot \beta_-}{N}, \label{p-}
 \end{eqnarray}
 and the correspondent Hamiltonian function is
 \begin{equation}
  \rm H=\rm \frac{Ne^{-3\Omega}}{24} \left[-P_\Omega^2 + P_+^2 + P_-^2 -48\Lambda  e^{6\Omega}
    +384\pi G M_\gamma e^{-3(\gamma-1)\Omega} \right]=0, \label{hami1}
 \end{equation}
 together with barotropic state equation $\rm p=\gamma \rho$, the Hamilton-Jacobi equation is obtained when we
substitute $\rm P_{q^\mu}\to \frac{dS_i}{dq^\mu}$ into (\ref{hami1}). In what follows, we should consider the gauge $N=1$.

\section{COMMUTATIVE CLASSICAL SOLUTIONS}
\subsection{Commutative Classical Solutions \'a la WKB}

The quantum Wheeler-DeWitt (WDW) equation for these models is  obtained by making the canonical quantization $\rm P_{q^\mu}$
by $\rm -i \partial_{q^\mu}$ in (\ref {hami1})
with $\rm q^\mu=(\Omega, \beta_+,\beta_-)$ is
\begin{equation}
  \rm \frac{e^{-3\Omega}}{24} \left[\frac{\partial^2}{\partial \Omega^2} - \frac{\partial^2}{\partial \beta_+^2} -
  \frac{\partial^2}{\partial \beta_-^2}  -\lambda  e^{6\Omega}
 +   b_\gamma e^{-3(\gamma-1)\Omega} \right]\Psi=0.
\label{wdw1}
\end{equation}
where $\rm \lambda=48 \Lambda$, $\rm b_\gamma=384\pi GM_\gamma$. We now proceed to apply the WKB semiclassical approximation using the ansatz
\begin{equation}
\rm \Psi\left(\Omega,\beta_\pm\right)=e^{i[S_1(\Omega)+S_2(\beta_+) + S_3(\beta_-)]},
\end{equation}
into (\ref{wdw1}), and without any loss of generality, one can consider the condition $\frac{d^2 S_i}{d q_i^2}$ be small i.e.,
\begin{equation}
\rm \Big(\frac{d S_1}{d \Omega}\Big)^{2}>> \frac{d^2 S_1}{d \Omega^2},\qquad
\Big(\frac{d S_2}{d \beta_+^2}\Big)^{2}>> \frac{d^2 S_2}{d \beta_+^2}, \qquad
 \Big(\frac{d S_2}{d \beta_-^2}\Big)^{2}>>\frac{d^2 S_2}{d \beta_-^2},
\end{equation}
to get the classical Einstein-Hamilton-Jacobi equation
\begin{equation}
\rm -\Big(\frac{d S_1}{d \Omega }\Big)^{2}+\Big(\frac{d S_2}{d \beta_+ }\Big)^{2}+ \Big(\frac{d S_3}{d \beta_- }\Big)^{2}
- \lambda e^{6\Omega}+ b e^{-3(\gamma-1) \Omega}=0,
\end{equation}
which can be separate in a set of differential equations
\begin{eqnarray}
\rm -\Big(\frac{d S_1}{d \Omega }\Big)^{2}+a_1^2- \lambda e^{6\Omega}+ b e^{-3(\gamma-1) \Omega} &=&\rm 0, \label{omega}\\
\rm \Big(\frac{d S_2}{d \beta_+ }\Big)^{2}&=&\rm  n_1^2, \label{+} \\
\rm \Big(\frac{d S_3}{d \beta_- }\Big)^{2}&=&\rm  p_1^2, \label{-}
\end{eqnarray}
where $\rm a_1^2$, $\rm n_1^2$ and $\rm p_1^2$ are the separation constants and their relations is $\rm a_1^2=n_1^2+p_1^2$. Therefore using the relations between
(\ref{p-}), (\ref{omega}), (\ref{+}) and (\ref{-})
we have the following equations of motion
\begin{eqnarray}
\rm \pm \sqrt{a_1^2- \lambda e^{6\Omega}+ b_\gamma e^{-3(\gamma-1) \Omega}  }& \equiv & -12 e^{3\Omega}\frac{\dot \Omega}{N},\label{oo}\\
\rm \pm n_1 &\equiv& 12 e^{3\Omega} \frac{\dot \beta_+}{N},\label{++}\\
\rm \pm p_1 &\equiv& 12 e^{3\Omega} \frac{\dot \beta_-}{N}.\label{--}
\end{eqnarray}
 The main master equation to solved in the gauge $\rm N=1$, is
\begin{equation}
\rm \frac{dt}{12}=\frac{d\Omega}{\sqrt{a_1^2 e^{-6\Omega}+ b_\gamma e^{-3(\gamma+1)\Omega}-\lambda}},
\end{equation}
the other two equations (\ref{++}) and (\ref{--}) are trivially integrable.
For  particular stadium of the universe evolution, given by the $\gamma$ parameter,
we present these classical solutions in table \ref{tabc1}.

\begin{center}
\begin{tabular}{|l| l|}\hline
 Case & Commutative solutions\\ \hline
$\gamma=-1,\ \ \Lambda \not=0,\ \ \rm \rho_{-1}=M_{-1}$&$\rm \Omega=  \frac{1}{3} \, Ln \, \left[ \frac{e^{2qt}-4a_1^2}{16qe^{qt}} \right]$,
\qquad $\rm \ \ \ \ \ q^2=24\pi G M_{-1} -3\Lambda$,\\
&$\rm \beta_+= \pm\frac{2}{3}\frac{n_1}{a_1} \, arctanh\, \left[\frac{e^{qt}}{2a_1} \right],\quad \rm a_1^2=n_1^2 + p_1^2 ,$\\
&$\rm \beta_-= \pm\frac{2}{3}\frac{p_1}{a_1} \, arctanh\, \left[\frac{e^{qt}}{2a_1} \right].$ 
\\ \hline  $\gamma=1,\ \ \Lambda <0,\ \ \rm \rho_1=M_1e^{-6\Omega}$
&$\rm \Omega=  \frac{1}{3} \, Ln \, \left[ \frac{e^{2qt}-4a_1^2}{16qe^{qt}} \right],$\qquad $\rm \qquad q= \sqrt{3|\Lambda|},$ \\
\ \ &$ \rm \beta_+=\pm  \frac{2}{3}\frac{n_1}{a_1} \, arctanh\, \left[\frac{e^{qt}}{2a_1} \right],$\ \ \ \ $\rm a_1^2=n_1^2 + p_1^2+384\pi G M_1,$\\
&$ \rm \beta_-=  \pm\frac{2}{3}\frac{p_1}{a_1} \, arctanh\, \left[\frac{e^{qt}}{2a_1} \right].$  
\\ \hline$\gamma=1,\ \ \Lambda =0,\ \  \rm \rho_1=M_1e^{-6\Omega}$&$\rm \Omega=\frac{1}{3}Ln\, [\frac{a_1}{4} t]$,
\qquad $\rm \qquad \qquad a_1^2=n_1^2 + p_1^2 +384\pi G M_1$, \\
&$\rm \beta_+=\pm Ln\,[t^{-\frac{n_1}{3a_1}}],$ \\
&$\rm \beta_-=\pm Ln\,[t^{-\frac{p_1}{3a_1}}].$ \\ \hline
$\gamma=0,\ \ \Lambda =0,\ \  \rm \rho_0=M_0e^{-3\Omega}$&$\rm \Omega=\frac{1}{3}Ln\, \left[ \frac{b_0 t^2}{64} +\frac{a_1 t}{4} \right]$,
\qquad $\rm \ \ \ b_0=384 \pi G M_0,$ \\
&$\rm \beta_+= \pm \frac{n_1}{3 a_1}\, Ln\,\left[\frac{16a_1 + b_0 t}{t} \right]$, \quad $\rm \ \ \ a_1^2=n_1^2+p_1^2,$
\\ &$\rm \beta_-=\pm \frac{p_1}{3 a_1}\, Ln\,\left[\frac{16a_1 + b_0 t}{t} \right].$\\ \hline
\end{tabular}
\emph{\label{tabc1} Table \ref{tabc1}. Classical Solutions for $\gamma=-1,1,0$,
and constraints $q$, $a_1$ and $b_0.$}
\end{center}

\subsection{Classical Solutions via Hamiltonian Formalism}
In order to find the commutative equation of motion, we use the classical phase space
variables $\rm (\Omega, \beta_\pm)$, where the Poisson algebra for these minisuperspace
variables are
\begin{equation}
\rm \left\{ \Omega,\beta_\pm  \right\}=\left\{\beta_+,\beta_-  \right\}=
\left\{P_\Omega,P_\pm  \right\}=\left\{P_+,P_-  \right\}=0, \qquad \left\{q^\mu,P_{q^\mu} \right\}=1,
\label{cbracket}
\end{equation}
and recalling the Hamiltonian equation (\ref{hami1}),
we obtain the classical solutions with the following procedure.

The classical equations of motion for the phase variables $\Omega$, $\beta_\pm$,  $P_\pm$, and $P_{\Omega}$ are
\begin{eqnarray}
\rm \dot{\Omega}&=&\rm \{\Omega, H\}=-\frac{1}{12} e^{-3\Omega}P_\Omega, \label{do1}\\
\rm \dot{\beta_-}&=&\rm \{\beta_-, H\}=\frac{1}{12} e^{-3\Omega}P_-,\label{beta-}\\
\rm \dot{\beta_+}&=&\rm \{\beta_+, H\}=\frac{1}{12} e^{-3\Omega} P_+,\label{beta+}\\
\rm \dot{P}_\Omega&=&\rm \{P_\Omega, H\}=\frac{1}{8}e^{-3\Omega}\left[-P^2_\Omega+ P_-^2 + P_+^2 +\lambda e^{6\Omega}+\gamma b_\gamma
e^{-3(\gamma-1)\Omega} \right] ,\label{gauge1}\\
\rm \dot{P}_-&=&\rm \{P_-, H\}=0,\qquad \to \qquad P_-=\pm p_1 =const. \label{p--}\\
\rm \dot{P}_+&=&\rm \{P_+, H\}=0,\qquad \to \qquad  P_+=\pm n_1 =const.\label{p++}
\label{classical}
\end{eqnarray}

Introducing (\ref{hami1}) into  (\ref{gauge1}), we have
\begin{equation}
\rm  8e^{-3\Omega}\dot{P}_\Omega=2\lambda+(\gamma -1)b_\gamma e^{-3(\gamma+1)\Omega},
\end{equation}
which can be integrate to obtain the relation for $\rm P_\Omega$
\begin{equation}
\rm P_\Omega= \pm \sqrt{a_1^2- \lambda e^{6\Omega}+ b_\gamma e^{-3(\gamma-1) \Omega}},
\label{ppomega}
\end{equation}
where $\rm a_1^2=n_1^2+p_1^2$.

The set of equations  (\ref{do1}), (\ref{beta-}) and
(\ref{beta+}) are equivalents to the set of equations (\ref{oo}),
(\ref{++}) and (\ref{--}), equations used to obtain the classical
solutions.

In summary, the anisotropic parameters are a crescent function of the time
(see table \ref{tabc1}), and  the solutions obtained with the Hamiltonian formalism and
the WKB-like procedure are equivalents in Classical General Relativity.

\section{NONCOMMUTATIVE SOLUTIONS}

Let us begin introducing the noncommutative deformation of the minisuperspace \cite{compean} in the WDW equation,
this time, between all the variables of the minisuperspace, assuming that $\Omega_{\text nc}$ and
$\beta_{\pm\text nc}$ obey the commutation relation

 \begin{equation}
\rm [\Omega_{\text nc},\beta_{-\text nc}]=i\theta_1, \qquad
[\Omega_{\text nc},\beta_{+\text nc}]=i\theta_2, \qquad
[\beta_{-\text nc},\beta_{+\text nc}]=i\theta_3. \label{combi}
\end{equation}
Instead of working directly with the physical variables $\Omega$ and $\beta_\pm$ we may achieve all the above
solutions by making use of the auxiliary canonical variables $\Omega_{\text{nc}}$ and $\beta_{\pm{\text nc}}$
defined as
\begin{eqnarray}
\rm \Omega_{nc} &\equiv &\rm \Omega -\frac{\theta_1}{2} P_{-}-\frac{\theta_2}{2} P_{+}, \label{ncomega}\\
\rm \beta_{-nc} &\equiv &\rm \beta_{-} + \frac{\theta_1}{2} P_{\Omega}-\frac{\theta_3}{2}P_{+},\label{ncbmenos}\\
\rm \beta_{+nc} &\equiv& \rm \beta_{+} + \frac{\theta_2}{2}
P_{\Omega}+\frac{\theta_3}{2} P_{-}.\label{ncbmas}
\end{eqnarray}
maintaining the usual commutation relations between the fields,
i.e., $\rm [q^{\mu},q^{\nu}]=0$. A shift generalization for the
commutative symplectic structure can be made it through the change
\begin{equation}
\rm q^{\mu } \equiv q^\mu_{nc} +\frac{1}{2} \theta^{\mu \nu} P_{\nu},
\end{equation}
where $\rm \theta^{\mu\nu}$ is an antisymmetric matrix, and the identifications
$P_{\Omega}=P_{\Omega\text{nc}}$ and $P_\pm=P_{\pm\text{nc}}$.
With this shift and the usual canonical quantization $\rm P_{q^\mu}\to
-i\partial_{q^\mu}$, we arrive to the noncommutative WDW equation
\begin{equation}
  \rm \left[\frac{\partial^2}{\partial \Omega_{nc}^2} -
\frac{\partial^2}{\partial \beta_{+nc}^2} -
  \frac{\partial^2}{\partial \beta_{-nc}^2}  -\lambda
e^{6\Omega_{nc} }+ b_\gamma e^{-3(\gamma-1)\Omega_{nc} } \right]\Psi(\Omega,\beta_{\pm})=0,
\label{ncwdw1}
\end{equation}
where $\rm \lambda=48 \Lambda$, $\rm b_\gamma=384\pi GM_\gamma$. At
this point we have a noncommutative WDW equation and noncommutative
hamiltonian. In what follows, we shall consider a wave function and
apply the WKB procedure to obtain classical solutions.

\subsection{Noncommutative Classical Solutions \'a la WKB}
In order to find noncommutative classical solutions through the WKB approximation, we use the fact that  $\rm e^{i\theta \frac{\partial}{\partial x}}
e^{\eta x}\equiv e^{i\eta \theta}e^{\eta x}$, and the ansatz for the
wavefunction $\rm \Psi(\Omega_{nc},\beta_{\pm nc})=
e^{i[S_1(\Omega_{nc}) \pm n_{1}\beta_{+nc} \pm p_{1}
\beta_{-nc}]}$, where we use explicitly  $\rm S_2(\beta_{+nc})=\pm
n_{1} \beta_{+nc}$ and  $\rm S_3(\beta_{-nc})=\pm p_{1} \beta_{-nc}$
to get the classical noncommutative Einstein-Hamilton-Jacobi (EHJ) equation
\begin{equation}
\rm -\Big(\frac{d S_1}{d \Omega_{nc}}\Big)^{2}+\Big(\frac{d S_2}{d \beta_{+nc} }\Big)^{2}+ \Big(\frac{d S_3}{d \beta_{-nc} }\Big)^{2}
- \lambda e^{6 \Omega_{nc}}+ b e^{-3(\gamma-1)  \Omega_{nc}}=0,
\end{equation}
which can be separate in a set of differential equations
with $ \rm m_{1}^2=n_{1}^2 + p_{1}^2$. We have the following noncommutative equations of motion
\begin{eqnarray}
\rm \pm \sqrt{a_1^2- \lambda e^{6 \Omega_{nc}}+ b_\gamma e^{-3(\gamma-1)  \Omega_{nc}}  }& \equiv & -12 e^{3 \Omega_{nc}}\frac{\dot \Omega_{nc}}{N},\label{oonc}\\
\rm \pm n_1 &\equiv& 12 e^{3 \Omega_{nc}} \frac{\dot \beta_{+nc}}{N},\label{++nc}\\
\rm \pm p_1 &\equiv& 12 e^{3 \Omega_{nc}} \frac{\dot \beta_{-nc}}{N}.\label{--nc}
\end{eqnarray}One just need to be careful in (\ref{oonc}), (\ref{++nc}) and (\ref{--nc}), and
apply the chain rule to the variables (\ref{ncomega}), (\ref{ncbmenos}) and (\ref{ncbmas}),
in order to get the right solution, $\dot \beta_{-nc}=\frac{\partial \beta_{-}}{\partial t}+
\frac{\partial \beta_{-nc}}{\partial P_{\Omega}}\frac{\partial p_{\Omega}}{\partial t}+
\frac{\partial \beta_{-nc}}{\partial P_{+}}\frac{\partial p_{+}}{\partial t}+
\frac{\partial \beta_{-nc}}{\partial P_{-}}\frac{\partial p_{-}}{\partial t}=\dot \beta_-+\frac{\theta_1 }{2} \dot P_{\Omega}$.  In this
sense, all solutions to find in the commutative case, remain for the
noncommutative case with the corresponding shift, as we show in the table \ref{tabnc1}.\\
\begin{center}
\begin{tabular}{|l| l|}\hline
 Case & Noncommutative Solutions\\ \hline
$\gamma=-1,\ \ \Lambda \not=0,\ \ \rm \rho_{-1}=M_{-1}$&$\rm \Omega_{nc}=\frac{1}{3} \, Ln \, \left[ \frac{e^{2qt}-4a_1^2}{16qe^{qt}}\right]
-\frac{\theta_1}{2}p_1-\frac{\theta_2}{2} n_1$,  \\
$\rm a_1^2=n_1^2 + p_1^2$ ,&$\rm \beta_{+nc}= \pm\frac{2}{3}\frac{n_1}{a_1} \, arctanh\, \left[\frac{e^{qt}}{2a_1} \right]+
\frac{\theta_2}{8}\left(\frac{e^{qt}}{4}+ a_1^2 e^{-q t} \right) -\frac{\theta_3}{2}p_1, $\\
$\rm q^2=24\pi G M_{-1} -3\Lambda$,&$\rm \beta_{-nc}= \pm\frac{2}{3}\frac{p_1}{a_1} \, arctanh\, \left[\frac{e^{qt}}{2a_1} \right]+
\frac{\theta_1}{8} \left(\frac{e^{qt}}{4}+ a_1^2 e^{-q t} \right)+\frac{\theta_3}{2}n_1,$ \\ \hline   
$\gamma=1,\ \ \Lambda <0,\ \ \rm \rho_1=M_1e^{-6\Omega}$ &
$\rm \Omega_{nc}=  \frac{1}{3} \, Ln \, \left[\frac{e^{2qt}-4a_1^2}{16qe^{qt}} \right]-\frac{\theta_1}{2}p_1-\frac{\theta_2}{2} n_1,$
\quad $\rm q= \sqrt{3|\Lambda|},$ \\
 $\rm a_1^2=n_1^2 + p_1^2+384\pi G M_1,$ &$ \rm \beta_{+nc}=\pm  \frac{2}{3}\frac{n_1}{a_1} \, arctanh\,
\left[\frac{e^{qt}}{2a_1} \right]+ \frac{\theta_2}{8}
\left(\frac{e^{qt}}{4}+ a_1^2 e^{-q t} \right)-\frac{\theta_3}{2}p_1,$\\
&$ \rm \beta_{-nc}=  \pm\frac{2}{3}\frac{p_1}{a_1} \, arctanh\, \left[\frac{e^{qt}}{2a_1} \right]+ \frac{\theta_1}{8}
\left(\frac{e^{qt}}{4}+ a_1^2 e^{-q t} \right)+\frac{\theta_3}{2}n_1,$  \\ \hline 
$\gamma=1,\ \ \Lambda =0,\ \  \rm \rho_1=M_1e^{-6\Omega}$&$\rm \Omega_{nc}=\frac{1}{3}Ln\, [\frac{a_1}{4} t]-\frac{\theta_1}{2}
p_1-\frac{\theta_2}{2} n_1$, \\
$\rm a_1^2=n_1^2 + p_1^2 +384\pi G M_1$,&$\rm \beta_{+nc}=\pm Ln\,[t^{-\frac{n_1}{3a_1}}]+ \frac{\theta_2}{2} a_1-\frac{\theta_3}{2}p_1,$ \\
&$\rm \beta_{-nc}=\pm Ln\,[t^{-\frac{p_1}{3a_1}}]+ \frac{\theta_1}{2} a_1+\frac{\theta_3}{2}n_1,$ \\ \hline 
$\gamma=0,\ \ \Lambda =0,\ \  \rm \rho_0=M_0e^{-3\Omega}$&$\rm \Omega_{nc}=\frac{1}{3}Ln\, \left[ \frac{b_0 t^2}{64} +\frac{a_1 t}{4}
\right]-\frac{\theta_1}{2} p_1-\frac{\theta_2}{2} n_1$, \\
$\rm b_0=384 \pi G M_0,$&$\rm \beta_{+nc}= \pm \frac{n_1}{3 a_1}\, Ln\,\left[\frac{16a_1 + b_0
t}{t} \right]+ \frac{\theta_2}{2} \sqrt{a_1^2 + \frac{b_0 t^2}{64} +\frac{a_1 t}{4}}-\frac{\theta_3}{2}p_1$,\\
$\rm a_1^2=n_1^2+p_1^2,$&$\rm \beta_{-nc}=\pm \frac{p_1}{3 a_1}\, Ln\,\left[\frac{16a_1 + b_0 t}{t} \right]+
\frac{\theta_1}{2} \sqrt{a_1^2 + \frac{b_0 t^2}{64} +\frac{a_1 t}{4}}+\frac{\theta_3}{2}n_1.$\\ \hline
\end{tabular}\\
\emph{\label{tabnc1} Table \ref{tabnc1}. Noncommutative solutions for,
$\gamma=-1,1,0,$  and constraints $q$, $a_1$
and $b_0$.}

\end{center}
\begin{figure}
\includegraphics[width=8cm]{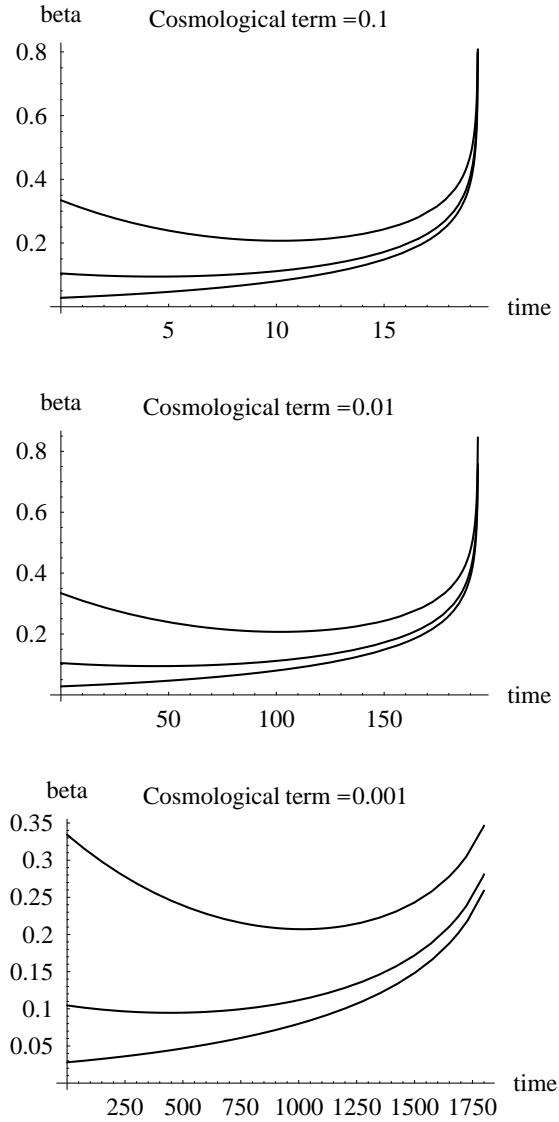}
\caption{\label{be1} Plots of $ \rm \beta_{\pm nc}$ that appear in the second line
in the table \ref{tabnc1},
 using the values in the parameters $\rm n_1=1, p_1=1, b_0=10$ and
$\theta=0,  0.05 , 0.2$, from bottom to top in the figure.
The possible isotropization is saw in function of the curvature, but it appears again in this fields. }
\end{figure}

In the following  plots we present as example, the second line in
the table \ref{tabnc1}, using different small values to the
cosmological constant, we can see that the anisotropic  parameters
$\beta_{\pm nc}$ for some particular value in the $\theta$
noncommutative parameter, tends to a constant curvature in some
range of the cosmological time. When this occur, the anisotropic
scale factors tend to anisotropic one, but next go to anisotropic
again. We  should have a process where this anisotropic behavior
does not appear again, for example, introducing in the model other
class of matter or energy, such as dark matter and dark energy. In
the noncommutative space,  it is possible to find one range in the
$\rm \theta$ parameter that produce the dynamical isotropization to
the model. But this isotropization does not occur in the other gauge
$\rm N=24 e^{3\Omega}$, when the $\rm \beta_{\pm nc}$, always is a
crescent function of the time (see table \ref{tabnc24}), independent
to the values in the $\theta$ parameters.

\subsection{Noncommutative Classical Solutions \'a la Hamilton}
Now the natural extension is to consider the noncommutative version
of our model, with the idea of noncommutative between the three
variables $\rm (\Omega_{nc},\beta_{\pm nc})$, for that purpose we
have two approaches the first one is, to modify the Poisson
structure, in this approach  the hamiltonian is not modify; in the
second approach we modify the hamiltonian via shift in the variables
but the symplectic structure stay intact. For the first case, as we
said we have the usual hamiltonian (\ref{hami1}), but the symplectic
structure is modify as follow
\begin{eqnarray}
\rm \left\{P_{\Omega},P_{\pm } \right\}_\star=\left\{P_{+},P_{-}
\right\}_\star=0, \qquad \left\{q^\mu,P_{q^\mu} \right\}_\star=1, \\
\rm \left\{\Omega,\beta_{-}\right\}_\star=\theta_1, \quad
\left\{\Omega,\beta_{+}  \right\}_\star=\theta_2, \quad
\left\{\beta_{-},\beta_{+} \right\}_\star=\theta_3.
\label{ncbracket}
\end{eqnarray}
where the $\star$ is the Moyal product \cite{Szabo2}. In the second case, the hamiltonian is modify by the shift
(\ref{ncomega}),(\ref{ncbmenos}) and (\ref{ncbmas}) resulting
\begin{equation}
\rm H_{nc}= \frac{Ne^{-3 \Omega_{nc}}}{24} \left[-P_{\Omega}^2 + P_{+}^2
+ P_{-}^2 -\lambda  e^{6 \Omega_{nc}}
    +b_\gamma e^{-3(\gamma-1) \Omega_{nc}} \right]=0,
 \label{non-hami11}
 \end{equation}
but the symplectic structure is the one that we know, the
commutative one (\ref{cbracket}).

The noncommutative equations of motion, for the first formalism that
we exposed have  the original variables, but with the modified
symplectic structure,
\begin{eqnarray}
\dot{q^\mu_{nc}}&=&\rm \{q^\mu, H\}_\star ,
\nonumber\\
\dot{P^\mu_{nc}}&=&\rm \{P^\mu, H\}_\star ,
\end{eqnarray} and for the second formalism we use the
shifted variables but with the original (commutative) symplectic
structure
\begin{eqnarray}
\dot{q^\mu_{nc}}&=&\rm \{q^\mu_{nc}, H_{nc}\} ,
\nonumber\\
\dot{P^\mu_{nc}}&=&\rm \{P^\mu_{nc}, H_{nc}\} ,
\end{eqnarray}
in  both approaches we have the same result. Therefore the equations of motion take the form
\begin{eqnarray}
\rm {\dot \Omega}_{nc}&=&\rm \{\Omega, H\}_\star=\{\Omega_{nc},
H_{nc}\}=-\frac{e^{-3 \Omega_{nc}}}{12} P_{\Omega},\\
\rm {\dot \beta}_{-nc}&=&\rm \{\beta_{-}, H\}_\star=\rm \{\beta_{-nc}, H_{nc}\}=\frac{e^{-3 \Omega_{nc}}}{12}P_{-}+\frac{\theta_1 }{2} \dot P_{\Omega}   ,\\
\rm {\dot \beta}_{+nc}&=& \rm \{\beta_{+}, H\}_\star=\{\beta_{+nc}, H_{nc}\}=\frac{e^{-3 \Omega_{nc}}}{12} P_{+} +\frac{\theta_2}{2} \dot P_{\Omega} ,\label{nc22} \\
\rm \dot{P}_\Omega&=&\rm \{P_{\Omega},H\}_\star=\rm
\{P_\Omega,H_{nc}\}=\frac{e^{-3 \Omega_{nc}}}{8}\left[6\lambda e^{6 \Omega_{nc}}+3(\gamma-1)
b_\gamma e^{-3(\gamma-1) \Omega_{nc}} \right],\label{ppomegac}\\
\rm \dot{P}_-&=&\rm \{P_{-}, H\}_\star=\rm \{P_-, H_{nc}\}=0, \quad \to \quad P_-=p_1,\label{nc3} \\
\rm \dot{P}_+&=&\rm \{P_+, H\}_\star=\rm \{P_+,H_{nc}\}=0, \quad \to \quad P_+=n_1. \label{nc4}
\label{classicalnc}
\end{eqnarray}
if we proceed as in the commutative case we get the solutions showed in the table IVA.

\section{Conclusions}
In this work we present the equivalence in General Relativity between the
WKB-approximation  and Hamilton formalism, in the commutative and noncommutative scenarios,
 this was achieved by means
 of a comparative study of the exact solutions for the Bianchi type I cosmological
 model coupled to  barotropic perfect fluid and cosmological term. As we can see
 the solution $\rm \Omega_{nc}$ is  the commutative solution  plus a function on
 $\theta_i$, independent of time.
 However, we have that in the physical gauge $\rm N=1$, the $\rm \beta_{\pm nc}$
 noncommutative solutions  suffers drastic changes with respect to the $\beta_\pm$
 commutative evolution.
 These changes  give the possibility that in some ranges on the parameter
 $\theta_i$ and cosmological constant, occurs a  dynamical isotropization,
 i.e.,  $\beta_{nc}\to \mbox{ a constant curvature}$. In other hand,
in the gauge $\rm N=24 e^{3\Omega}$ (see appendix A and B), in all  cases considered
the influence of the noncommutativity is
encoded  as an addition smooth function on time, to the classical solutions
and the change is
qualitatively very remarkable only for certain ranges on the $\rm \theta_i$
parameters, but in general, the anisotropization is not modified, in sense
 commutativity is recovered dynamically.
Besides we show that the definitions of the noncommutative commutators can be
applied to all the variables of the minisuperspace. This approach can be used for other
 Bianchi cosmological model, which will be reported elsewhere.

\acknowledgments{This work was supported in part by CONACyT grant 47641 and
Promep grant UGTO-CA-3. Many calculations where done by Symbolic Program REDUCE 3.8. }

\appendix{}

\section{Commutative classical solutions in the Gauge $ N=24 e^{3\Omega}$}

\subsection{Commutative Classical Solutions \'a la WKB}
The master equation becomes
\begin{equation}
\rm 2 dt=\frac{d\Omega}{\sqrt{a_1^2-\lambda e^{6\Omega}+ b_\gamma e^{-3(\gamma-1)\Omega}}},
\end{equation}
and the other two equations are immediately integrable. Again for
particular cases in the $\gamma$ parameter, we present the classical solutions, table \ref{tabcom24}\\

\begin{center}
\begin{tabular}{|l| l|}\hline
 Case & Commutative solutions\\ \hline
$\gamma=-1,\ \ \Lambda \not=0,\ \ \rm \rho_{-1}=M_{-1}$&$\rm \Omega= \frac{1}{6}Ln\, \left[ -\frac{a_1^2}{384\pi G M_{-1} -48\Lambda}\, Sech^2 \left(6 a_1 t \right)\right],$\quad
 $\rm a_1^2=n_1^2 + p_1^2,$\\
&$\rm \beta_+= \pm2 n_1 t,$\\
&$\rm \beta_-= \pm 2 p_1 t.$ \\ \hline 
 $\gamma=1,\ \ \Lambda \not = 0,\ \ \rm \rho_1=M_1e^{-6\Omega}$
&$\rm \Omega=\frac{1}{6}\, Ln \, \left[\frac{a_1^2}{48\Lambda}\, Sech^2(6 a_1 t) \right] ,$\ \ \ \ $\rm a_1^2=n_1^2 + p_1^2 +384\pi G M_1,$\\
\ \ &$ \rm \beta_+= \pm2 n_1 t,$ \\&$ \rm \beta_-= \pm2 p_1 t.$ \\ \hline    
$\gamma=1,\ \ \Lambda =0,\ \  \rm \rho_1=M_1e^{-6\Omega}$&$\rm \Omega=2 \sqrt{a_1^2 + b_1} t,$ \qquad
 $\rm \qquad \qquad a_1^2=n_1^2 + p_1^2+384\pi G M_1,$\\
&$\rm \beta_+=-2 n_1 t,$ \\
&$\rm \beta_-=-2 p_1 t.$ \\ \hline
$\gamma=0,\ \ \Lambda =0,\ \  \rm \rho_0=M_0e^{-3\Omega}$&$\rm \Omega=\frac{1}{3}Ln\, \left[-\frac{a_1^2}{b_0} \, sech^2(3a_1 t)\right] ,$\quad
 $\rm \ \ b_0=384 \pi G M_0,$ \\
&$\rm \beta_+= \pm2n_1 t, $ \qquad $\rm a_1^2=n_1^2+p_1^2,$ \\
&$\rm \beta_-=\pm2 p_1 t.$\\ \hline
$\gamma=\frac{1}{3},\ \ \Lambda =0,\ \  \rm \rho_0=M_{\frac{1}{3}}e^{-4\Omega}$ & $ \rm \Omega=\frac{1}{2}Ln\,
\left[-\frac{a_1^2}{b_{\frac{1}{3}}} \, sech^2(2a_1 t)\right],$ \quad  $\rm a_1^2=n_1^2+p_1^2,$\\
&$\rm \beta_+= \pm2n_1 t, $ \qquad $\rm \qquad \qquad \qquad b_{\frac{1}{3}}=384 \pi G M_{\frac{1}{3}}$,\\
&$\rm \beta_-=\pm2 p_1 t.$\\ \hline
\end{tabular}\\
\emph{\label{tabcom24} Table \ref{tabcom24}. Classical Solutions for $\gamma=-1,\frac{1}{3},1,0,$  and constraints  $a_1,$ $b_0$ and $b_1.$} \label{iB}
\end{center}

\subsection{Classical Solutions via Hamiltonian formalism}
With the gauge fixed to  $\rm N=24 e^{3\Omega}$ we can see that the hamiltonian takes the form
\begin{equation}
\rm H= -P_\Omega^2 + P_+^2 + P_-^2 -\lambda  e^{6\Omega}
    +b_\gamma e^{-3(\gamma-1)\Omega} =0.
 \label{hami-n}
 \end{equation}
The Poisson brackets structure yields to equations of motion
\begin{eqnarray}
\rm \dot{\Omega}&=&\rm \{\Omega, H\}=-2 P_\Omega, \label{do}\\
\rm \dot{\beta_-}&=&\rm \{\beta_-, H\}=2P_-,\qquad \to \qquad \beta_-=\pm 2p_1 t,\\
\rm \dot{\beta_+}&=&\rm \{\beta_+, H\}=2P_+,\qquad \to \qquad \beta_+=\pm 2n_1 t,\\
\rm \dot{P}_\Omega&=&\rm \{P_\Omega, H\}=\left[+6\lambda e^{6\Omega}+3(\gamma-1) b_\gamma e^{-3(\gamma-1)\Omega} \right] ,\label{gaugen}\\
\rm \dot{P}_-&=&\rm \{P_-, H\}=0,\qquad \ \ \  \to \qquad P_-=\pm p_1 =const. \label{nn1}\\
\rm \dot{P}_+&=&\rm \{P_+, H\}=0,\qquad \ \ \ \to \qquad  P_+=\pm n_1 =const. \label{pp1}
\label{cclassical}
\end{eqnarray}
Using (\ref{hami-n}), introducing (\ref{nn1}) and (\ref{pp1}), we obtain the expression for $\rm P_\Omega$
\begin{equation}
\rm P_\Omega= \sqrt{m_1^2 -\lambda  e^{6\Omega} +b_\gamma e^{-3(\gamma-1)\Omega} },
\end{equation}
being self-consistent with equation (\ref{gaugen}), where $\rm a_1^2=n_1^2+p_1^2$. Introducing this equation into (\ref{do}) we get
the master equation found  to  solve the Einstein field equation in this gauge, where the classical solutions
are presented in table IIB.

\section{Noncommutative Classical Solutions}
\subsection{Noncommutative Classical Solutions in the Gauge $\rm N=24 e^{3\Omega}$ \'a la WKB and via Hamiltonian formalism}
The noncommutative solutions in the space
$\rm q^{\mu }$ become

\begin{center}
\begin{tabular}{|l| l|}\hline
 Case & Noncommutative Solutions\\ \hline
$\gamma=-1,\ \ \Lambda \not=0,\ \ \rm \rho_{-1}=M_{-1}$&$\rm \Omega_{nc}=
\frac{1}{6}Ln\, \left[ -\frac{a_1^2}{384\pi G M_{-1} -48\Lambda}\,
Sech^2 \left(6 a_1 t \right)\right]-\frac{\theta_1}{2}
p_1-\frac{\theta_2}{2} n_1,$\\
  $\rm a_1^2=n_1^2 + p_1^2,$&$\rm \beta_{+nc}= \pm2 n_1 t+ \frac{\theta_2 a_1}{2}\, tanh(6a_1 t)-\frac{\theta_3}{2}p_1,$,\\
&$\rm \beta_{-nc}= \pm 2 p_1 t+ \frac{\theta_1 a_1}{2}\, tanh(6a_1 t)+\frac{\theta_3}{2}n_1,$  \\ \hline  
 $\gamma=1,\ \ \Lambda \not = 0, \rm \rho_1=M_1e^{-6\Omega}$
&$\rm \Omega_{nc}=\frac{1}{6}\, Ln \, \left[\frac{a_1^2}{48\Lambda}\, Sech^2(6 a_1 t) \right]-\frac{\theta_1}{2}
p_1-\frac{\theta_2}{2} n_1 ,$\\
  $\rm a_1^2=n_1^2 + p_1^2 +384\pi G M_1,$ &$ \rm \beta_{+nc}= \pm2 n_1 t+ \frac{\theta_2 a_1}{2}\, tanh(6 a_1 t)-\frac{\theta_3}{2}p_1,$ \\
&$ \rm \beta_{-nc}= \pm2 p_1 t+ \frac{\theta_1 a_1}{2} \, tanh(6 a_1 t)+\frac{\theta_3}{2}n_1,$\\ \hline   
$\gamma=1,\ \ \Lambda =0,\ \  \rm \rho_1=M_1e^{-6\Omega}$&$\rm \Omega_{nc}=2 a_1 t-\frac{\theta_1}{2} p_1-\frac{\theta_2}{2} n_1,$ \\
  $\rm a_1^2=n_1^2 + p_1^2+384\pi G M_1,$ &$\rm \beta_{+nc}=-2 n_1 t+ \frac{\theta_2}{2} a_1-\frac{\theta_3}{2}p_1,$ \\
&$\rm \beta_{-nc}=-2 p_1 t+ \frac{\theta_1}{2} a_1+\frac{\theta_3}{2}n_1,$ \\ \hline   
$\gamma=0,\ \ \Lambda =0,\ \  \rm \rho_0=M_0e^{-3\Omega}$&$\rm \Omega_{nc}=\frac{1}{3}Ln\, \left[-\frac{a_1^2}{b_0} \, sech^2(3a_1
t)\right] -\frac{\theta_1}{2}
p_1-\frac{\theta_2}{2} n_1,$  \\
$\rm b_0=384 \pi G M_0,$&$\rm \beta_{+nc}= \pm2n_1 t + \frac{\theta_2 a_1}{2}
\, tanh(3a_1 t)-\frac{\theta_3}{2}p_1,$ \\
$\rm a_1^2=n_1^2+p_1^2,$&$\rm \beta_{nc}-=\pm2 p_1 t+ \frac{\theta_1 a_1}{2} \, tanh(3a_1 t)+\frac{\theta_3}{2}n_1.$\\ \hline  
$\gamma=\frac{1}{3},\ \ \Lambda =0,\ \  \rm \rho_0=M_{\frac{1}{3}}e^{-4\Omega}$ & $ \rm \Omega_{nc}=\frac{1}{2}Ln\,
\left[-\frac{a_1^2}{b_{\frac{1}{3}}} \, sech^2(2a_1 t)\right]-\frac{\theta_1}{2} p_1-\frac{\theta_2}{2} n_1,$ \\
 $\rm a_1^2=n_1^2+p_1^2,$ &$\rm \beta_{+nc}= \pm2n_1 t + \frac{\theta_2 a_1}{2}\, tanh(2 a_1 t) -\frac{\theta_3}{2}p_1$ \\
 &$\rm \beta_{-nc}=\pm2 p_1 t+ \frac{\theta_1 a_1}{2}\, tanh(2 a_1 t) +\frac{\theta_3}{2}n_1.$\\ \hline
\end{tabular}\\
\emph{\label{tabnc24} Table \ref{tabnc24} Noncommutative  solutions for
$\gamma=-1,\frac{1}{3},1,0,$
and constraints $a_1,$ $b_0$ and $b_1.$}
\label{iiB}
\end{center}

\end{document}